\documentclass{article}
\usepackage{epsfig}
\usepackage{amsmath}
\usepackage{amssymb}
\usepackage[hang,nooneline,scriptsize]{subfigure}

\usepackage[dvips]{color}
\usepackage[normalem]{ulem}

\newcommand{\ee}{\end{equation}}
\newcommand{\eea}{\end{eqnarray}}
\newcommand{\be}{\begin{equation}}
\newcommand{\bea}{\begin{eqnarray}}

\begin{document}

\title{\LARGE \bf Proca Q Balls and their Coupling to Gravity}
  \author{
  \large  Y. Brihaye$^a$, Th. Delplace$^a$ $\:${\em and}$\:$
  Y. Verbin$^b$ \thanks{Electronic addresses: yves.brihaye@umons.ac.be; thomas.delplace@student.umons.ac.be; verbin@openu.ac.il } }
 \date{ }
   \maketitle
    \centerline{$^a$ \em Physique Th\'eorique et Math\'ematiques, Universit\'e de Mons,}
   \centerline{\em Place du Parc, B-7000  Mons, Belgique}
     \vskip 0.4cm
   \centerline{$^b$ \em Department of Natural Sciences, The Open University
   of Israel,}
   \centerline{\em Raanana 43107, Israel}

 \maketitle
\begin{abstract}
Extending the Proca Lagrangian of a massive complex-valued vector field by self-interaction potential,
we construct a large class  of spherically symmetric solutions in flat Minkowski background as well as in the self-gravitating case. Our solutions  encompass Proca Q-balls and Proca stars known in the literature, but present new features and go beyond the limited region of parameter space charted so far. A special emphasis is set to the domain of existence
of the solutions in relation with the coupling constants of the potential and to the
critical phenomena limiting this domain.
\end{abstract}

\maketitle

\medskip
 \ \ \ PACS Numbers: 04.70.-s,  04.50.Gh, 11.25.Tq

\section{Introduction}\label{Introduction}
\setcounter{equation}{0}

Vector field analogues of Q-balls, boson stars and Q-stars as well as their cylindrically-symmetric versions have been studied quite extensively in the recent couple of years. Some works that can be regarded as representative of the various directions of investigation are Loginov's \cite{Loginov2015} which studies non-topological solitons produced in flat space by a self-interacting (complex) vector field, Brito et al. \cite{BritoEtAl2015} which constructs self-gravitating spherical solutions of the complex (Proca) vector field with a mass term only (i.e. Proca Stars) and Landea and Garcia \cite{Landea-Garcia2016} who promote the global U(1) symmetry to be local.

These studies and others \cite{Duarte-Brito2016,sanchis2017,bv2017}
are motivated by suggestions of massive spin-1 particles as a dark matter ingredient \cite{Holdom1985,ArkaniHamedEtAl2008,Pospelov-Ritz2008,GoodsellEtAl2009} and by the interest in generalizing the scalar non-topological solitons \cite{FriedbergLeeSirlin1976,Coleman1985,Lynn1988,Jetzer1989,Lee-Pang1991} to systems of vector fields and study the similarities and differences.

However, the interest in massive vector fields is not new, but can be traced to the 1930s with the original work of Proca \cite{Proca1936,Proca1937,Poenaru2005}. There has been also some action during the years since - see e.g. \cite{Obukhov1999,Vuille2002}.

Another aspect of these theories that was studied quite extensively is black hole solutions that were found to overcome the no-vector-hair theorems \cite{Bekenstein1972a,Bekenstein1972b} exhibiting a vector hair in either the Abelian \cite{HerdeiroEtAl2016,Fan2016} or non-Abelian case \cite{Ponglertsakul-Winstanley2016}.

The dynamics of these vector systems is determined by the following action (we stay Abelian):
\be \label{totalAction}
S = \int d^4 x \sqrt{- g} \left (\frac{R}{16 \pi G} -\frac{1}{4} F^*_{\mu\nu}F^{\mu\nu} - U(A^*_\mu A^\mu) \right )
\ee
where  $A_{\mu}$ is the complex vector potential and $F_{\mu\nu}=\partial_{\mu}A_{\nu}-\partial_{\nu}A_{\mu}$ is the corresponding field strength. $R$ is Ricci scalar and $G$ Newton's constant; we use the Landau-Lifshitz sign conventions with a ``mostly minus'' metric.

The potential function may contain only a mass term as in the original Proca theory, or be a higher order polynomial like
\be \label{potential}
U(\psi) = -\frac{m^2}{2} \psi - \frac{\lambda}{4} \psi^2 - \frac{\nu}{6} \psi^3  \,\,\,\,\,\, , \,\,\,\psi=A^*_\mu A^\mu
\ee

The higher order polynomial may support localized solutions even in flat space without invoking gravity \cite{Loginov2015} unlike the ``pure'' Proca system. At any rate, the potential function breaks explicitly the local $U(1)\times U(1)$ symmetry of the pure (double) Maxwell theory (the kinetic term), and leaves a global $U(1)$ only.

In this paper we perform a detailed analysis of the spherically-symmetric solutions of the  self-interacting vector system either self-gravitating or in flat background and extend the limits to further regions in parameter space uncharted before. We find (perhaps surprisingly) that localized stable solutions exist in certain conditions for both signs of the two coupling constants of the potential, $\lambda$ and $\nu$.

This phenomenon is related to the fact that unlike the scalar case, the potential $U(A^*_\mu A^\mu)$ is unbounded from below because of the indefinite spacetime norm. Indeed, the energy (or mass) density turns out still to be bounded from below for a certain range of parameters, but stable well-behaved localized solutions exist much outside this region of positivity. Actually, this should not cause any worry because the potential can be ``corrected'' by a higher order positive definite power of $\psi$ which will make it bounded from below without any noticeable change of the localized solutions which are also limited to a finite interval around the origin of field space.

Moreover, this same reasoning may be repeated in the scalar case in order to construct non-topological solitons
as solutions for unbounded potentials with the proviso that this is only an approximation to a full potential which is bounded from below (see e.g. \cite{Mielke2002-6}).

In order to understand the general structure of the solutions, we use the fact that for a given potential (i.e. a given set of parameters $m$, $\lambda$, and $\nu$) the spherically symmetric solutions are essentially a one parameter family characterized by the central value of the vector field (or another equivalent parameter). This fact enables us to analyze these families of solutions, to
calculate their main characteristics like global charge (interpreted as particle number) and mass and study their variation. We use the mass to charge ratio in order to study the stability of these solutions against ``fission'' into a number of smaller stable (vector) Q-balls or Q-stars. We identify regions of stability and find that the structure is much richer than in the scalar case.

We will start the discussion in the next section by presenting the model, the field equations and the asymptotic behavior of the solutions. Next we solve (numerically) the field equations for Q-balls and study them in flat background. Finally we analyze the role of gravity on these solutions. We will still use the term ``Proca stars'' for these solutions, although our solitons are always ``made of'' self-interacting vector fields rather than the ``pure'' Proca field.

\section{Model, Ansatz and Field Equations}
\label{The Model}
\setcounter{equation}{0}

The field equations for the self-interacting vector field as derived from (\ref{totalAction}) are
\be \label{ProcaEq}
\nabla_\mu F^{\mu\nu} - 2\frac{dU}{d\psi} A^\nu = 0
\ee
which are supplemented with the constraint (analogous to the Lorentz condition for the Maxwell field\footnote{the analogy is not perfect since this condition is not just a gauge fixing as in the Maxwell theory.})
\be \label{ProcaConstraint}
\nabla_\mu \left( \frac{dU}{d\psi} A^\mu \right)= 0
\ee
The conserved global $U(1)$ current is
\be \label{ProcaCurrent}
J^\mu = -\frac{i}{2}\left( F^{*\mu\nu}A_\nu - F^{\mu\nu}A^*_\nu \right)
\ee

If gravity is ``switched on'', one should solve these equations in a self-consistent way with the Einstein equations written in terms of the Einstein tensor $G_{\mu\nu}$
\be \label{EinsteinEqG}
G_{\mu\nu} + 8\pi G T_{\mu\nu}= 0
\ee
where the energy-momentum tensor is given by
\be \label{energy-mom-tensor}
T_{\mu\nu} = \frac{1}{2}\left(F^{*\lambda}_{\mu}F_{\lambda\nu} + F^{*\lambda}_{\nu}F_{\lambda\mu}\right) + \frac{1}{4} F^*_{\kappa\lambda}F^{\kappa\lambda}g_{\mu\nu} - \frac{dU}{d\psi}\left(A^*_\mu A_\nu + A^*_\nu A_\mu  \right) + U(\psi) g_{\mu\nu}
\ee

We assume a spherically symmetric  form of metric given by the line element
\be \label{spher-metric}
ds^2=g_{\mu\nu}dx^\mu dx^\nu= A^2(r) N(r) dt^2- \frac{1}{N(r)} dr^2- r^2 d \Omega_2^2
\ee
while for the vector field we assume the radial ``electric'' configuration
\be \label{cyl-VectorField}
A_\mu dx^\mu = e^{-i\omega t}\left( a_0 (r) dt + i a_1(r) dr  \right)
.\ee
The two components $a_0 (r)$ and $a_1 (r)$ are assumed to be real.
They satisfy the ``Lorentz'' condition which takes now the form
\be \label{Lorentzspher}
\omega (m^2 + \lambda \psi + \nu \psi^2) a_0 +
\frac{A N}{r^2} \frac{d}{dr} [A N r^2 (m^2 + \lambda \psi + \nu \psi^2)a_1]=0
\ee
where now $\psi$ takes the form $\psi=a_0^2 /(A^2 N) - N a_1^2 $.

The ansatz (\ref{cyl-VectorField})
was first used  in \cite{Loginov2015} and \cite{BritoEtAl2015} following a similar one  \cite{friedberg:1976} in the context of non abelian gauge theories.
It takes advantage of the complex character of the Proca field.
The occurrence of the radial component $a_1$ -which can be gauged
away in the case of the Maxwell theory- is essential to support the soliton.
In particular the dependance of the equation on the frequency $\omega$ would vanish with $a_1=0$ or the
field $a_1$ would become non-dynamical while setting $\omega=0$.

The field equations (\ref{ProcaEq}) become
\begin{eqnarray}\label{ProcaEqspher-0}
 \omega \left( a'_0 - \omega a_1 \right) + N A^2 \left( m^2 + \lambda\psi + \nu \psi^2 \right)a_1 =0 \\
 \label{ProcaEqspher-1}
 \frac{A N}{r^2}\left[\frac{r^2}{A} \left( a'_0 - \omega a_1 \right) \right]' - \left( m^2 + \lambda\psi + \nu \psi^2 \right)a_0 =0
\end{eqnarray}
and it is easy to see that substituting (\ref{ProcaEqspher-0}) into (\ref{ProcaEqspher-1}) yields (\ref{Lorentzspher}). Alternatively, Eq. (\ref{ProcaEqspher-0}) is a linear combination of (\ref{Lorentzspher}) and (\ref{ProcaEqspher-1}).
These field equations may be derived from an effective Lagrangian which is obtained by substituting the ansatz (\ref{spher-metric})-(\ref{cyl-VectorField}) into the vector sector of the Lagrangian density in (\ref{totalAction}), namely:
\be \label{Leff}
L_{eff}=A r^2 \left(\frac{\left( a'_0 - \omega a_1 \right)^2}{2 A^2}
+\left(\frac{m^2}{2} \psi+ \frac{\lambda}{4} \psi^2 + \frac{\nu}{6} \psi^3\right)\right)
\ee
In order to obtain a boundary value problem, we set  the system in a form where
the equation for $a_1$ and $a_0$   are respectively of the first and second order.

The Einstein equations depend on the energy momentum tensor $T_\mu^\nu $ which turns out to be diagonal
for the static spherical case; the components are given by
\begin{eqnarray}\label{Tmunu}
T_0^0=\frac{\left( a'_0 - \omega a_1 \right)^2}{2 A^2}
+\left(m^2 + \lambda\psi + \nu \psi^2 \right)\frac{a_0^2}{N A^2}
-\left(\frac{m^2}{2} \psi+ \frac{\lambda}{4} \psi^2 + \frac{\nu}{6} \psi^3\right) \\
T_r^r=\frac{\left( a'_0 - \omega a_1 \right)^2}{2 A^2}
-\left(m^2 + \lambda\psi + \nu \psi^2 \right) {N a_1^2}
-\left(\frac{m^2}{2} \psi+ \frac{\lambda}{4} \psi^2 + \frac{\nu}{6} \psi^3\right) \\
T_\varphi^\varphi=T_\theta^\theta=-\frac{\left( a'_0 - \omega a_1 \right)^2}{2A^2}
-\left(\frac{m^2}{2} \psi+ \frac{\lambda}{4} \psi^2 + \frac{\nu}{6} \psi^3\right)
\end{eqnarray}
Substituting the spherically symmetric ansatz into the Einstein equations leads to two first order equations
for the metric fields $N(r)$ and $A(r)$:
\begin{eqnarray}
\label{gravity_eq}
     {\cal M}' &=& 4 \pi G r^2 T_0^0  , \ \ \\
     A' &=& 2 \pi G r A \left(\frac{a_0^2}{N^2 A^2} + a_1^2 \right) \left(m^2 + \lambda \psi + \nu \psi^2 \right)
\end{eqnarray}
where we write $N(r) = 1 - 2  {\cal M}(r)/r$.  The third Einstein equation is not independent, but  is a consequence of these equations.

\subsection{Physical parameters}
\label{PhysPar}
An important characteristic of the solutions is the global $U(1)$ charge.
It is readily obtained from the time component of the conserved current (\ref{ProcaCurrent}):
\be \label{ProcaCharge}
Q= - 4 \pi \int_0^\infty dr  \frac{r^2}{A} \left( a'_0 - \omega a_1 \right)a_1 =
\frac{4 \pi }{\omega} \int_0^\infty dr r^2 N A \left( m^2 + \lambda\psi + \nu \psi^2 \right)a_1^2
\ee
where the second expression was obtained by using (\ref{ProcaEqspher-0}). We assume of course that the integral converges for the localized solution we are after. Without loss of generality we will take $\omega>0$, so $Q>0$ as well. We remind the reader that this $U(1)$ charge has no relation to electromagnetism,  nor does the complex vector field that we study here. The self-interaction potential explicitly breaks gauge invariance and the vector field is not assumed to couple (except gravitationally) to other matter - e.g. additional scalar fields.   This is the reason why the gauge coupling constant ($e$) is absent from this paper. The global $U(1)$ charge $Q$ defined in (\ref{ProcaCharge}) is therefore interpreted as the total number of elementary vector (Proca) particles.

The solutions can further be characterized by their mass. We will use the
ADM (or gravitational) mass which reads from the asymptotic decay of the metric~:
$N(r) = 1 - 2 G M /r + o(1/r^2)$, i.e. using (\ref{gravity_eq})
\be \label{InertialMass}
M  = 4 \pi \int_0^\infty dr r^2  T_0^0
\ee
Note that the contribution from the potential term to the mass density (i.e. the two last terms of $T_0^0$) is not always positive definite, but it is so for $\lambda^2<4\nu m^2$ \cite{Loginov2015}.
All masses  that we calculated turn out to be  positive also outside this range.

One important aspect of the mass and the global charge together is in the  ratio $M/|Q|m$ which needs to be less than one in order for the solutions to be stable against ``fission'' into smaller Q-balls or Q-stars.
The condition $M/|Q|m<1$ is however not sufficient to guarantee the stability
under linear perturbation \cite{BritoEtAl2015} .

\subsection{Remarks on the potential}
\label{RemPot}
For the purpose of the discussion it is useful to mention some features
of the effective potential $V_{eff}(a_0,a_1)$ obtained by setting all fields to constants in the effective Lagrangian (\ref{Leff}) (and in the absence of dynamical gravity). The relevant potential takes the form~:
\be
\label{veff}
-V_{eff}(a_0,a_1) = \frac{m^2}{2} a_0^2 + \frac{\omega^2 - m^2}{2} a_1^2
+ \frac{\lambda}{4}(a_0^2 - a_1^2)^2
 + \frac{\nu}{6}(a_0^2 - a_1^2)^3
\ee
Assuming $\omega^2 \leq m^2$ (this is the case for all solutions obtained), the following statements follow~:
(i) The origin $a_0=a_1=0$ always constitutes a saddle point. (ii) In the case $\nu=0$, the configurations
$a_0=0, a_1 = \pm (m^2-\omega^2)/ \lambda$ constitute two local maxima. (iii) In the case $\nu > 0$
the configurations $a_0 = 0, a_1 = \pm a_{1,m}$ and $a_0 = 0, a_1 = \pm a_{1,s}$ with
\be
       a_{1,m}^2 = \frac{1}{2 \nu} ( \lambda - \sqrt{\lambda^2 + 4 \nu (\omega^2-m^2)}) \ \ , \ \
       a_{1,s}^2 = \frac{1}{2 \nu} ( \lambda + \sqrt{\lambda^2 + 4 \nu (\omega^2-m^2)})
\label{saddlepoint}
\ee
constitute respectively two local maxima and two saddle points of the potential.

\subsection{Rescaling}
\label{Rescaling}
The field equations depend a priori on the 3 parameters in the potential $m,\lambda,\nu$ and Newton's constant $G$. The equations can be written in a dimensionless form which reveals a dependence on two independent parameters only: We replace $r$ by $x=mr$ and scale the components $a_0$ and $a_1$ by a factor $\mu=m/\sqrt {|\lambda |}$ where $m$ is the mass of the vector field. Then the equations depend on the two dimensionless parameters
\be
     \overline \nu = \frac{\nu m^2}{\lambda^2} \ \ , \ \ \alpha = 4 \pi G \frac{m^2}{|\lambda|} = \frac{1}{2|\lambda|}\left(\frac{m}{m_{Pl}}\right)^2
     \label{RescPar}
\ee
where $m_{Pl}$ is the Planck mass.  In performing this rescaling we have assumed $\lambda \neq 0$. The reasons for this is that (i) we want
to study first the solutions in flat Minkowski background (the so called probe limit $\alpha =0$) and
(ii) we  found no localized solution exist in flat space with a mass term only. This suggests that a more-than-quadratic term
should be present in the potential. We therefore decided to impose the most economic quartic term to be present
allowing to treat $\nu$ as an extra coupling constant. The fact that vector Q-ball solutions exist with a quartic term only may be considered a sharp difference with respect to the scalar case, but actually we find the difference milder if we recall that  a consistent non-topological soliton may be obtained as a solution with a scalar potential which is unbounded from below, if it is also ``corrected'' far enough in field space by a higher order power of the field without any noticeable change of the solutions \cite{Mielke2002-6}.

Note that the above rescaling further implies that the frequency is
 rescaled according to $\overline \omega = \omega/m$; in the following, we will
present the data in the  dimensionless variables and omit the 'bar' of $\overline \omega$ and $\overline \nu$.

The dimensionless version of the vector field equations (\ref{Lorentzspher})-(\ref{ProcaEqspher-1}) are obtained formally by taking $m=1$ and $\lambda=1$ (and in the self-gravitating case also $\lambda=-1$) and replacing $r$ by the dimensionless radial variable $x$. For convenience, we give the rescaled equations in the appendix. We also comment that the rescaled version of the charge and mass are obtained by the same way exactly. They are related to one another by $\bar{M}=M m/\mu^2=M |\lambda|/m$ and $\bar{Q}=Qm^2/\mu^2=Q|\lambda|$. The stability ratio may be calculated both ways: $M/|Q|m = \bar{M}/\bar{|Q|}$.

\section{Q-balls~: Asymptotic form and boundary conditions}
\label{General}
\setcounter{equation}{0}
In the case $\alpha = 0$ (the probe limit), the relevant equations reduce to a system of coupled equations
of first and second order respectively for the  fields $a_1$ and $a_0$.
The regularity of the solutions at the center $x=0$ imposes $a_1(0)=0$ and $a'_0(0)=0$.

Inspecting the possible asymptotic behavior of the fields, we found two possibilities. The first is:
\be
\label{type_1}
       a_0 (x)    \propto \frac{e^{- \sqrt{1-\omega^2} \hspace{0.075cm} x}}{ x} \ \ , \ \
       a_1 (x)= - \frac{\omega}{1 - \omega^2} \frac{d a_0}{dx}
\ee
The solutions of this type will be referred to as Type-1. The form
\be
\label{type_0}
         a_0(x) = - \frac{A}{\omega x} + o(1/x^2) \ \ , \ \
          a_1(x) = A  - \frac{\omega}{1 - \omega^2} \frac{d a_0}{dx}
\ee
is also possible, provided the condition  $\omega^2 +  \lambda A^2 - \nu A^4=1$
is obeyed by the asymptotic constant $A$ and the frequency $\omega$.
 We will refer to these solutions as being Type-0.

In both cases, the asymptotic form implies $a_0(\infty)=0$.
Together with the regularity conditions $a_0'(0)=0, a_1(0)=0$
  this specifies the boundary value problem.
With a given choice
of the self-interacting potential (actually, by choosing (the rescaled) $\nu$ only),
the system admits solutions
for a limited interval of values  of the frequency $\omega$ or, equivalently,
 of the central field value $a_0(0)$ (these quantities are related through the equations).
The family of solutions labeled by $\omega$ then constitutes a branch of Q-balls. The first problem
is  to determine the pattern of solutions for the different choices of the coupling constant $\nu$.

It turns out that the asymptotic forms (\ref{type_1}),(\ref{type_0})
are both consistent with the  boundary conditions and regular solutions of both types indeed exist.
\begin{itemize}
\item {\bf Type-0} solutions have
 $a_1(\infty) \neq 0$ and $a'_1(\infty) = 0$. They have no finite mass
as seen by looking at (\ref{Tmunu}). However they are useful for the understanding of the pattern of localized solutions.
\item {\bf Type-1}  solutions  have
 $a_1(\infty) = 0$ and $a'_1(\infty) = 0$. These solutions have a finite mass
constituting Proca Q-balls.
\end{itemize}
As  discussed in the next section,
the occurrence (eventually the  co-existence) of  solutions of type-0 and type-1 depends
on the coupling constants and on the frequency $\omega$.

\section{Q-Ball Solutions in Flat Space}
\label{QBFlat}

As stated above, all coupling constants can then be scaled and for a given set of coupling constants the Q-balls  form a family parametrized by the frequency $\omega$ or, equivalently, by the value $a_0(0)$. In flat space we are left with a potential function which depends on a single parameter, the rescaled $\nu$.

\subsection{ Case $\nu = 0$}
\label{nu0}
As mentioned already, we have found that unlike previous assumptions in the literature, there exist Q-ball solutions with $\nu = 0$ i.e. the potential is a sum of the mass term and the quartic term only. This difference motivated us to check the case of $\lambda \leq 0$, but we found no Q-ball solutions in this case.
Let us thus first discuss the minimal case of $\nu = 0$ and $\lambda > 0$.
\begin{figure}[b!]
\begin{center}
{\includegraphics[width=8cm]{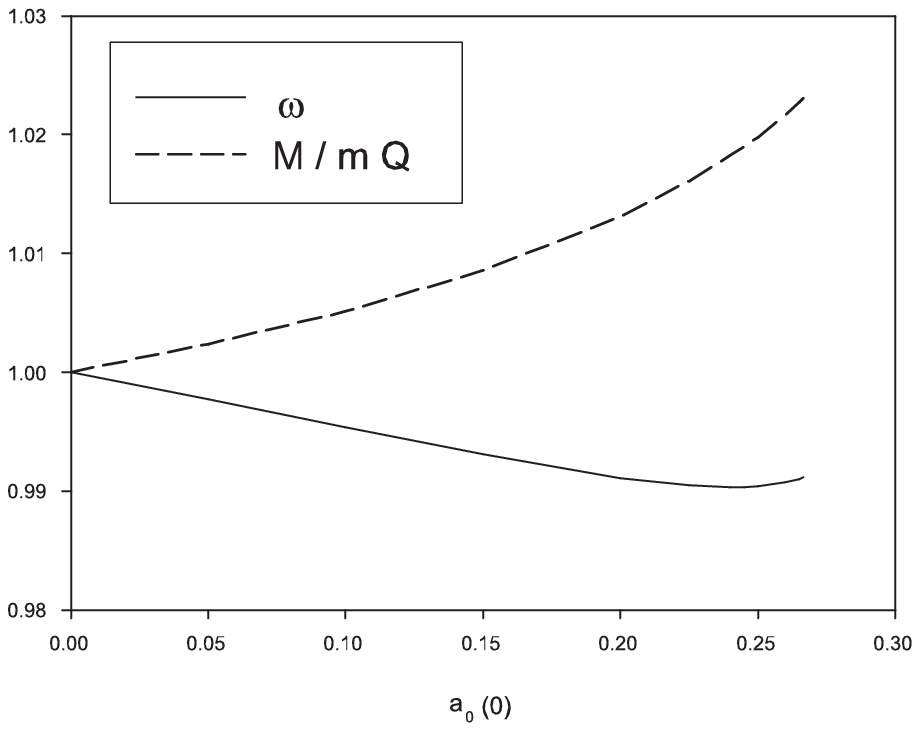}}
{\includegraphics[width=8cm]{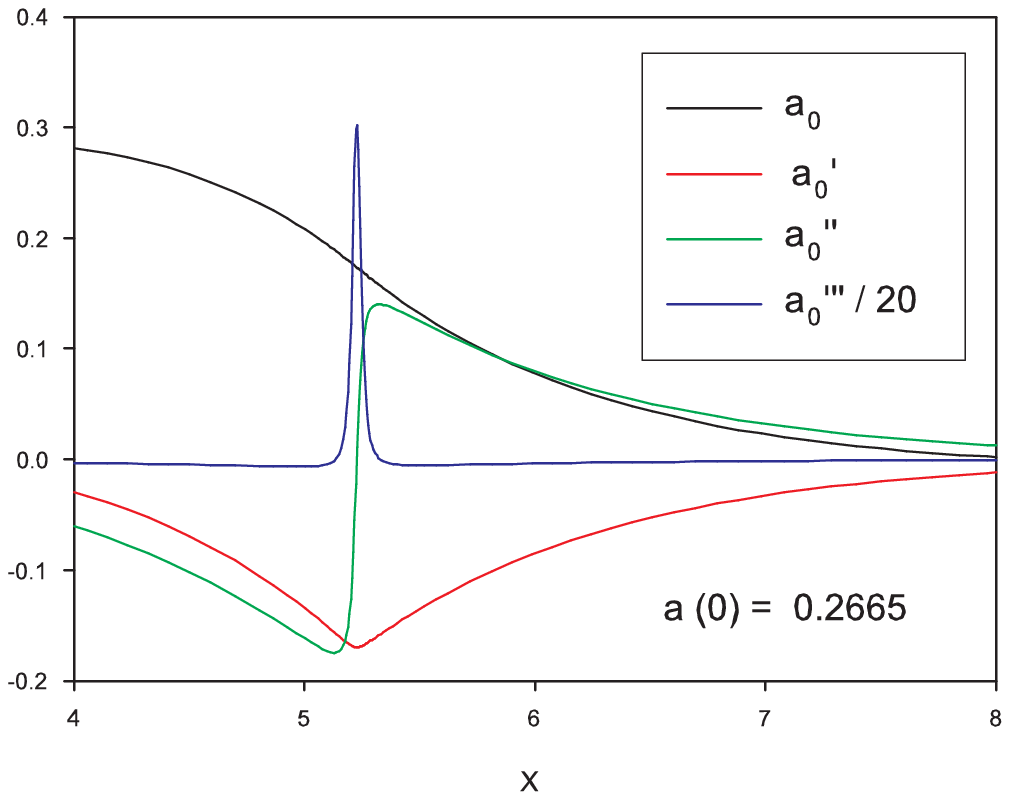}}
\caption{Left: $a_0(0)$-dependence of $\omega$ and the ratio $M/mQ$
for the solutions with $\lambda = 1$, $\nu=0$.
Right: Profiles of the solution
with $a_0(0) = 0.2665$ (corresponding to $\omega \approx 0.991$) near the branch termination point.
\label{case_nu_0}
}
\end{center}
\end{figure}

Forcing $\omega < 1$ in the equations, a family of
regular solutions can be constructed obeying the boundary conditions of type-1 solutions.
They possess a finite mass $M$ and charge $Q$. In the limit $\omega \to 1$, the Proca field tends to zero because the
functions $a_{0}$, $a_1$  both approach  the null function (in the sense that $|a_{k,max} \to 0$,  $k=0,1$).
However the convergence of the functions to zero is weak, as a result both the mass and the charge do not
approach zero for $\omega \to 1$ (note: this feature also holds for scalar Q-balls).
The relation between the frequency and $a_0(0)$ is shown by the solid line on Fig.\ref{case_nu_0}-Left.
When the  value $a_0(0)$ increases, the fields approach a configuration which seems to become
singular at a maximal value of $a_0(0)$.
In particular, the maximal value of the {\it third} derivative $a_0'''$ becomes infinite at
some intermediate radius $x_c$ (with $0 < x_c < \infty$).
The profiles of $a_0$ and  its derivatives are shown on Fig.\ref{case_nu_0}-Right
for the solution with $a_0(0) = 0.2665$, i.e. close to the maximal value.  For larger values of $a_0(0)$
the numerical integration  becomes difficult and strongly indicates
that the branch terminates into a singular configuration.
For all solutions we find for the stability ratio $M/mQ > 1$ (see Fig.\ref{case_nu_0}-Left).
 The vector Q-balls available with a purely quartic potential (plus mass term) are therefore classically unstable.

\subsection{ Case $\nu > 0$}
\label{nu>0}
\begin{figure}[t!]
\begin{center}
{\includegraphics[width=5cm, angle = 270]{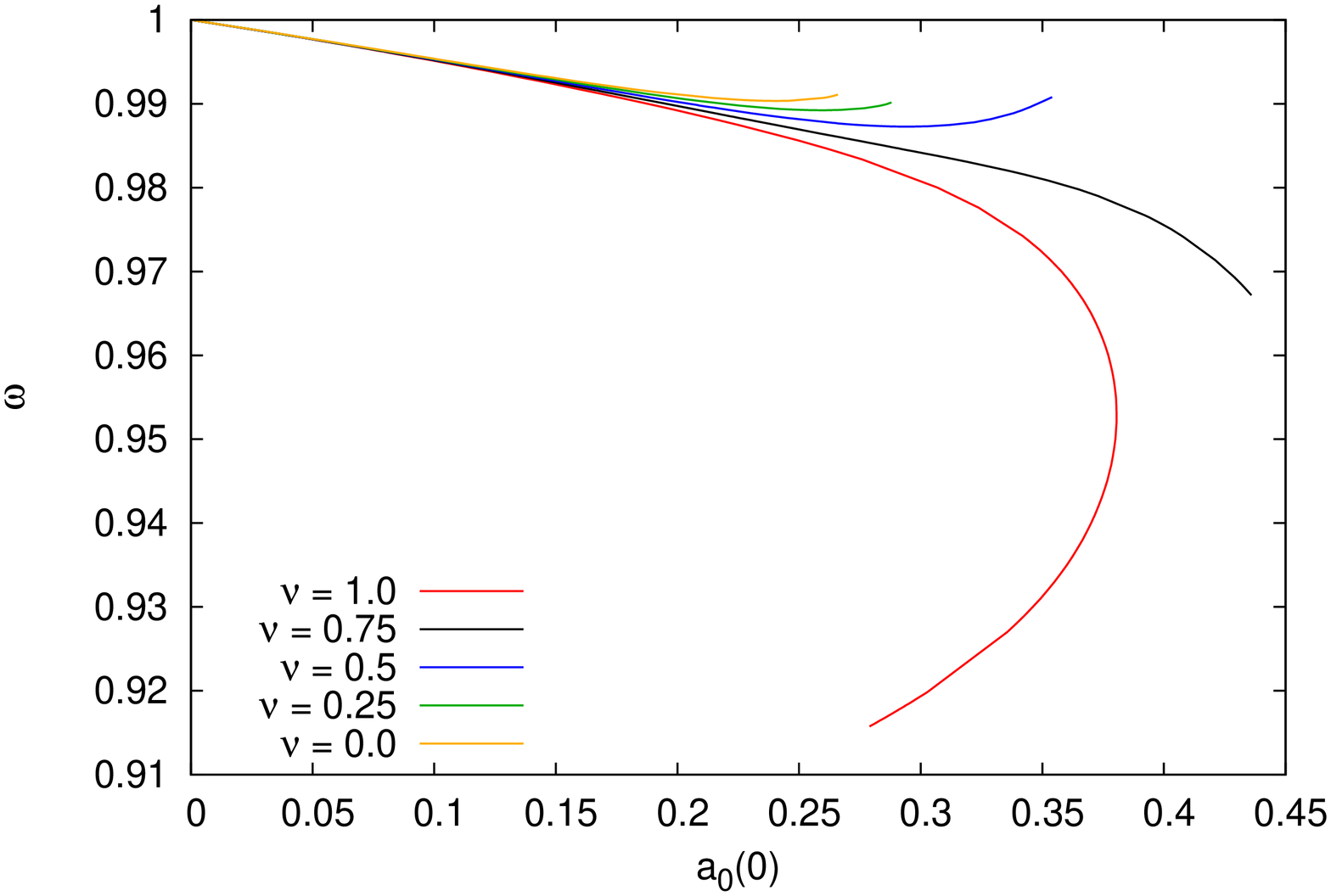}}
{\includegraphics[width=5cm, angle = 270]{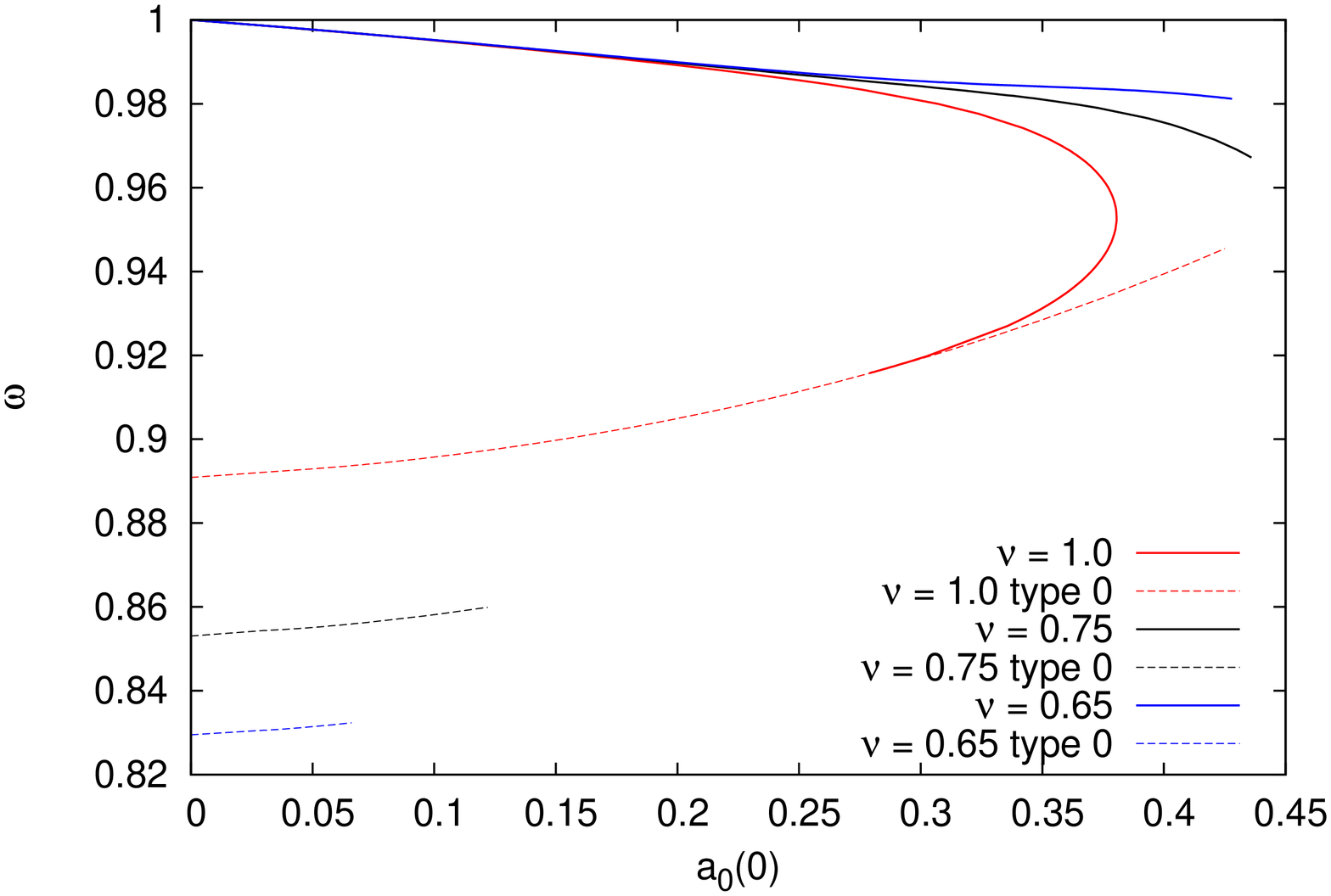}}
\caption{Left: Type-1 solutions. Dependence of the frequency $\omega$ on $a_0(0)$
for several values of $\nu$.
Right: Type-0 and Type-1 solutions. $\omega$ - $a_0(0)$ curves for three values of $\nu$.
\label{w_a0}
}
\end{center}
\end{figure}
\begin{figure}[b!]
\begin{center}
{\includegraphics[width=6cm, angle = 270]{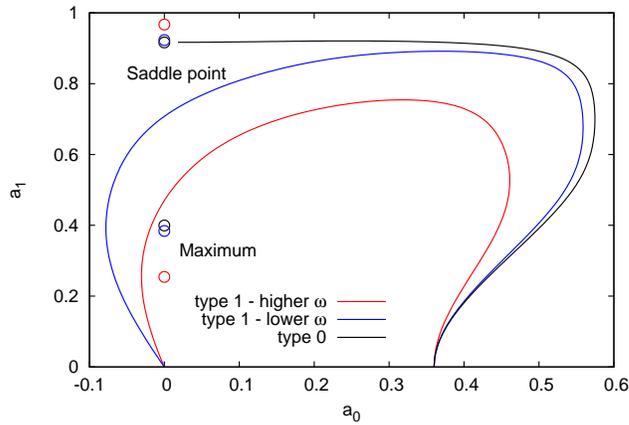}}
\caption{Trajectories in the ($a_0,a_1$) plane of the three solutions starting with $a_0(0) = 0.36$. The bullets represent the positions of the critical points of $V_{eff}$ with $\nu=1$. We didn't mark the saddle point at the origin.
\label{traj_nu=1}
}
\end{center}
\end{figure}
In Loginov's paper\cite{Loginov2015}  type-1 solutions  were studied for a specific choice of parameters which in our language translates to $\nu=1$ only. We now emphasize generic values of $\nu$
and show that the pattern of solutions is quite sensitive to $\nu$. The pattern is closely related to
emergence of second saddle point  of the effective potential as discussed in subsection \ref{RemPot}.

The type-1 solutions can be obtained by deforming progressively
the $\nu = 0$ solutions discussed above. The evolution of the domain of existence in terms of the parameter $a_0(0)$ is illustrated in  Fig. \ref{w_a0}-Left where the $\omega-a_0(0)$ relation is shown for
several values of $\nu$. It turns out that, when $\nu$ becomes sufficiently large (typically $\nu > 0.9$)
two type-1 solutions exist with the {\it same} value of $a_0(0)$ although distinguished by
two different values of the frequency $\omega$.
For the solution with the smaller frequency, the field $a_1(r)$ has a tendency to remain
constant  on an interval of $x$, forming a plateau before decreasing to its null asymptotic value.
The value of the function  $a_1(x)$ at the plateau is very close the
the value $a_{1,s}$ of the second saddle point (\ref{saddlepoint}) of the potential. Also the mass of these solutions become very large while $\omega$ decreases.

Independently of the existence of type-1 solutions, families of type-0 solutions can be constructed as well
when the parameter $\nu$ is sufficiently large. The numerical results show that type-0 solutions start to exist for for $\nu \simeq 0.6$ and exist  for a larger and larger
interval of the parameter $a_0(0)$ while increasing  $\nu$. This is
illustrated by  Fig.\ref{w_a0}-Right where the $\omega-a_0(0)$ relation is shown on
for $\nu= 0.65$, $0.75$ , $1.0$  (see the dashed lines on Fig.\ref{w_a0}-Right).

For $\nu > 0.9$ it turns out that the branches of type-0
and type-1 solutions join at some critical value $\omega_c$.
In the case $\nu =1.0$, the two curves in red in  Fig.\ref{w_a0} show the type-0 solution
(dashed lines) and type-1 (solid lines)
joining  for $\omega = \omega_c = 0.916$ , $a_0(0) = 0.279$.
In other words, this suggests  that the Type-1 branch of finite mass solutions bifurcates from the Type-0 branch.

It is tempting to visualize the different solutions as trajectories of a two-dimensional
motion of a particle in the effective potential (\ref{veff}) and  undergoing  friction
due to the derivative terms.
 The influence of the various critical points on the three trajectories corresponding to
 $\nu=1$ and $a_0(0) = 0.36$ can be appreciated in Fig. \ref{traj_nu=1}. The three solutions
 are presented in different colors and the positions of the critical points of the effective potential
 are indicated by the bullets. We see clearly that the trajectories ``prefer'' to end (asymptotically) at a saddle point and each type has its own preference. On their way to the saddle point at the origin the two type-1 solutions avoid the potential maximum and go around it.

\begin{figure}[t!!!]
\begin{center}
{\includegraphics[width=8cm, angle = 270]{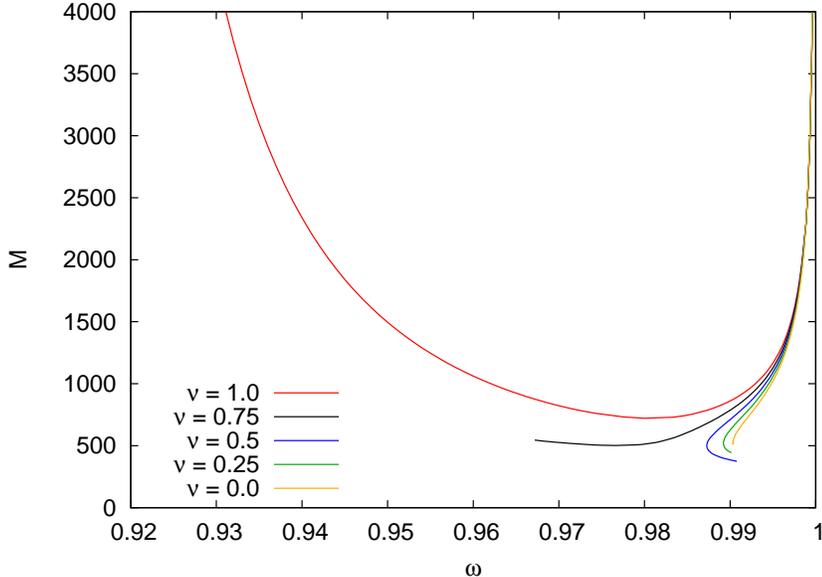}}
\caption{Dependence of the mass $M$ (in units of $m/|\lambda|$) on $\omega$  for several values of $\nu$.
\label{m_w}
}
\end{center}
\end{figure}

The $\omega$-dependence of the masses $M$ for several values of $\nu$
is shown in Fig. \ref{m_w}. Confirming our interpretation  of a bifurcation of the
type-1 branch  from the (divergent mass) type-0 branch, we observe that the mass of the solution
corresponding to $\nu = 1$ diverges in the limit $\omega \to \omega_c = 0.919$. The actual numerical values of the masses that we calculate are easily obtained from our dimensionless results. As discussed above, $\bar{M}$ measures the mass in units of $m/|\lambda|$ as
$\bar{M}=M |\lambda|/m$. So for any choice of the pair of $m$ and $|\lambda|$, the actual mass
can be read off Fig. \ref{m_w}.

\begin{figure}[t!]
\begin{center}
{\includegraphics[width=5cm, angle = 270]{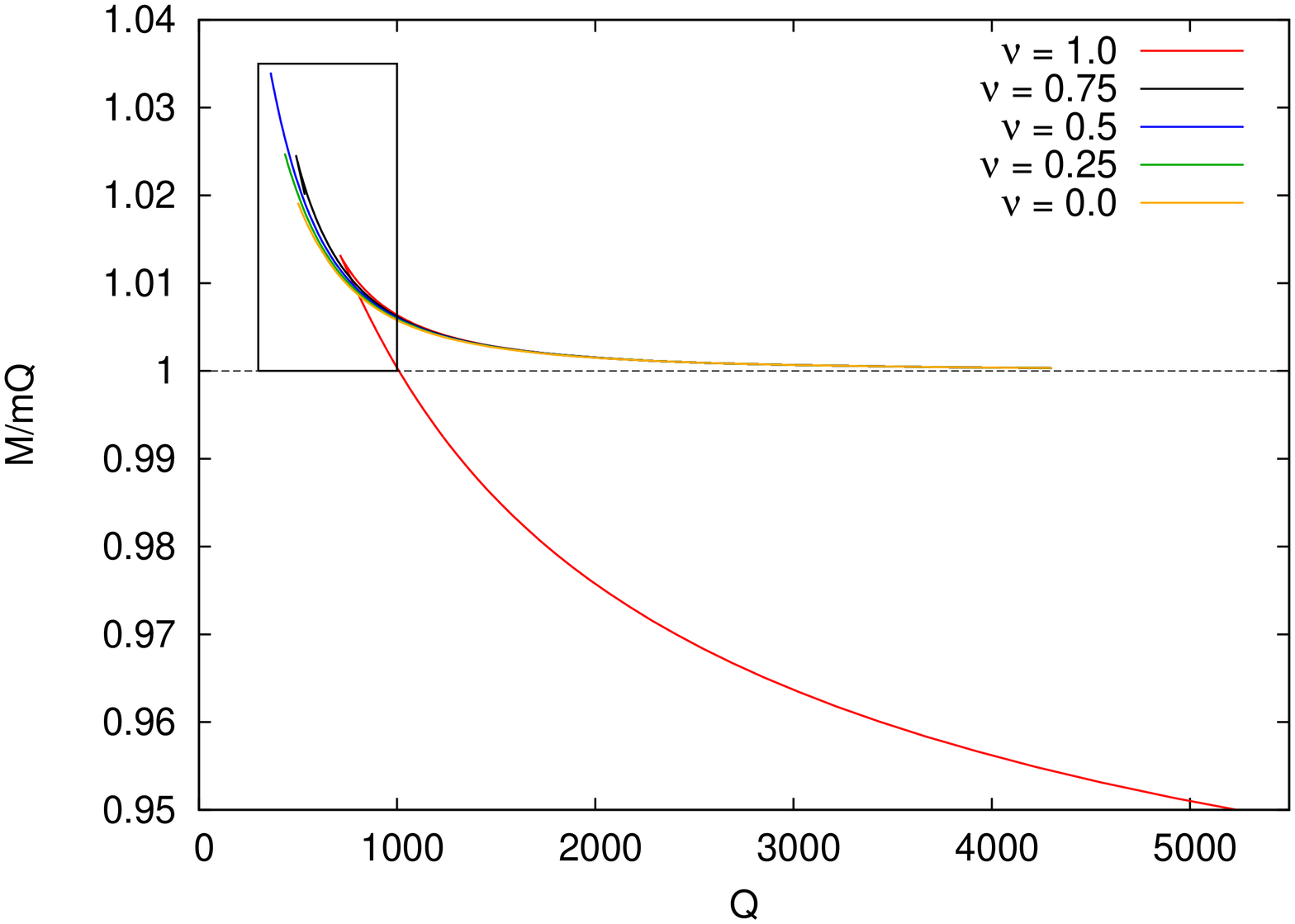}}
{\includegraphics[width=5cm, angle = 270]{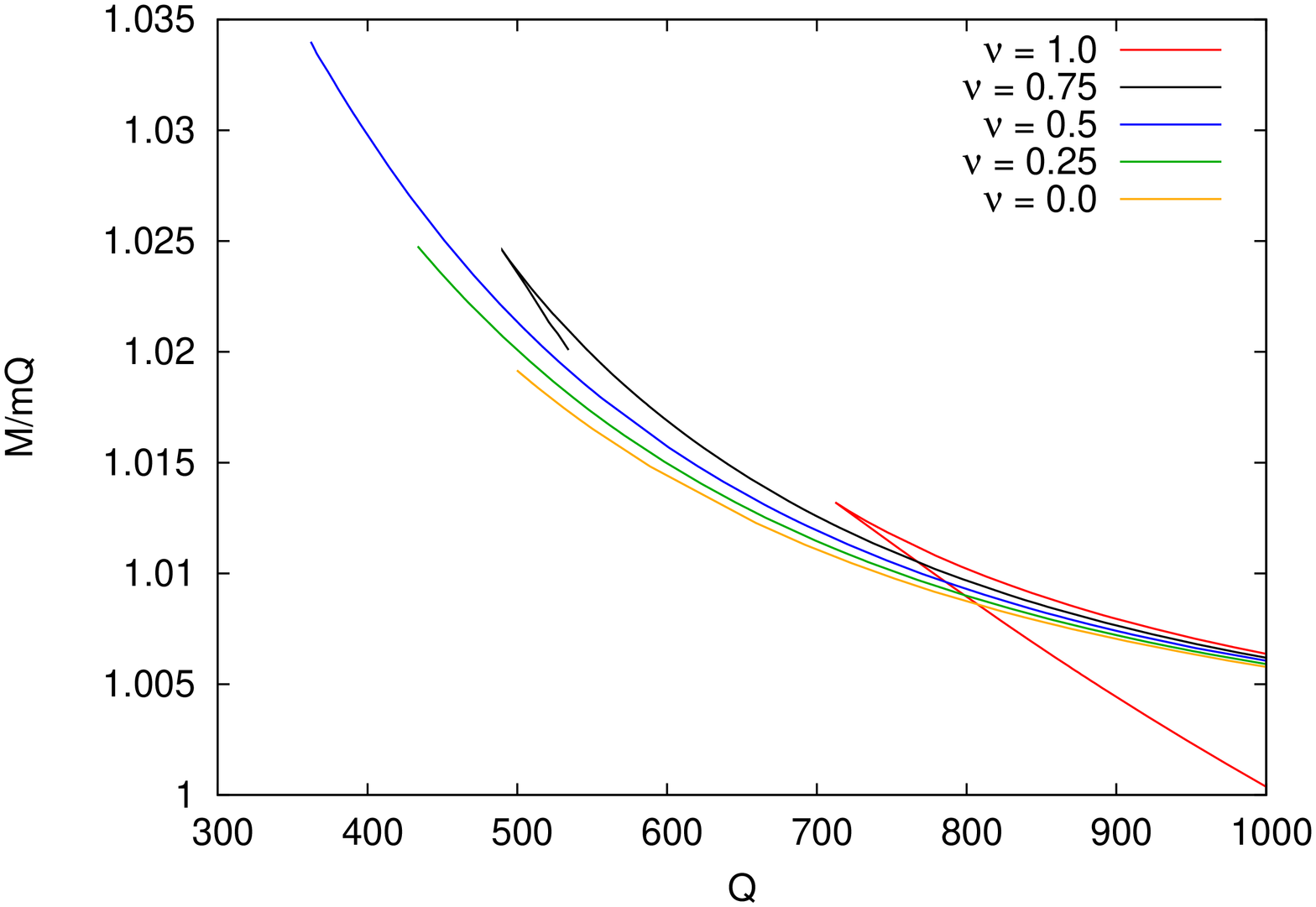}}
\caption{Left: $Q$-Dependence of the ratio $M/mQ$ for several values of $\nu$. The region in the box is zoomed in the second part of the figure.
Right: Zoom of the small $Q$ region marked on the left part.
\label{M_Q_Q}
}
\end{center}
\end{figure}

As far as the classical stability is concerned, we notice that the ratio $M/mQ$
becomes smaller than 1 only for large enough values of $\nu$ which allows for smaller frequencies and larger $Q$-values. For $\nu=1$ for example, the stability region consists of the segment of $\omega \leq  0.96$ and $Q\geq 1000$. When plotted against the particle number $Q$ in Fig. \ref{M_Q_Q},
the ratio $M/mQ$ presents a single smooth line (without spikes) for small $\nu$ and two segments divided by a spike for large $\nu$ (typically $\nu > 0.70$). A portion of the lower segment has $M/mQ < 1$ and corresponds to classically stable solutions.

\begin{figure}[b!]
\begin{center}
{\includegraphics[width=6cm, angle = 270]{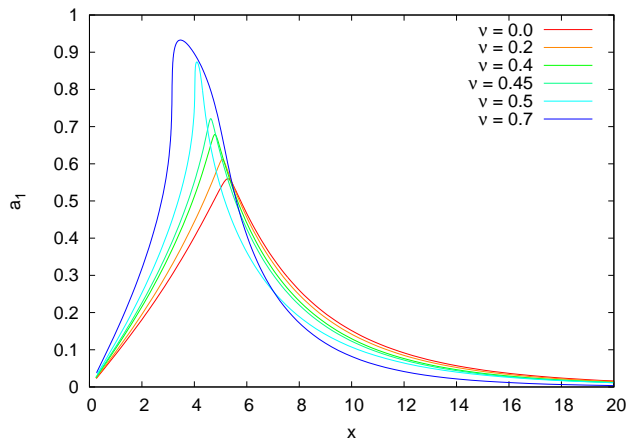}}
\caption{Profiles of the field $a_1$ close to the critical limit for several values of $\nu$.
\label{superposition}
}
\end{center}
\end{figure}

The reason why several  curves in Figs. \ref{w_a0} and \ref{m_w}
stop at specific values of the frequency, say $\omega = \omega_s$
is related to the fact that, when the frequency becomes too small,
the solutions approach configurations where the fields become singular at a particular radius $x_c$.
This was seen already on Fig. \ref{case_nu_0} for $\nu = 0$.
The profiles of the field $a_1(x)$ close to the critical limit and for several values of $\nu$
are superposed on Fig.\ref{superposition}. Three distinct behaviors are clearly observed
at a critical radius $x_c$~:
\begin{itemize}
\item For $0 \leq \nu \leq 0.45$, the type-1 branch stops because $a'''(x_c)$ tends to infinity for $\omega \to \omega_s$.
\item For $0.45 \leq \nu \leq 0.9$, the type-1 branch stops because $a'(x_c)$ tends to infinity for $\omega \to \omega_s$.
\item For $0.9 \leq \nu$, the type-1 branch stops because it bifurcates into the type-0 branch for $\omega \to \omega_s$.
\end{itemize}
For all the values of $\nu$ that we studied, the type-0
branch (when it exists) also tends to a singular configuration
when the parameter $a_0(0)$ approaches a maximal value.
In particular the field $a_0(x)$ approaches a wall-shape for  a finite radius $x_m$ before reaching
its non vanishing asymptotic value.
The numerical results indeed reveal that both $|a_1'(x_w)|$ and   $|a_0''(x_w)|$
tend to infinity.

\section{Proca Q-Stars}
We now discuss the influence of gravity on the Proca Q-balls constructed in the previous section.
For this purpose, we solved the field equations for $\alpha >0$.  For brevity,
we present the results for the case $\nu=0$ only, but we checked that many features are qualitatively  similar for $\nu > 0$.
Considering solutions with  generic values $\alpha >0$ and $\lambda >0$
is somehow redundant because solutions with $\alpha >0$ and $\lambda = 1$ are equivalent
after a suitable rescaling of the field and parameters to solutions with $\tilde \alpha=1$ and $\tilde \lambda = 1/ \alpha$ (see also Eq. (\ref{RescPar})).
Accordingly, the Proca stars of Brito et al. \cite{BritoEtAl2015} are recovered from our results by taking
the limit $\alpha \to \infty$.
However, using continuity arguments, gravitating solutions with $\lambda <0$ can reasonably be expected
to exist. We indeed produced such solutions
and showed that they do not survive in the probe limit $\alpha \to 0$.
The cases $\lambda = 1$ and $\lambda = -1$ are discussed separately
in the following.

Another result of adding gravity is the elimination of the type-0 solutions: i.e. they exist for $\alpha=0$ only.

\subsection{The case $\lambda = 1$}
\begin{figure}[t!]
\begin{center}
{\includegraphics[width=5cm, angle = 270]{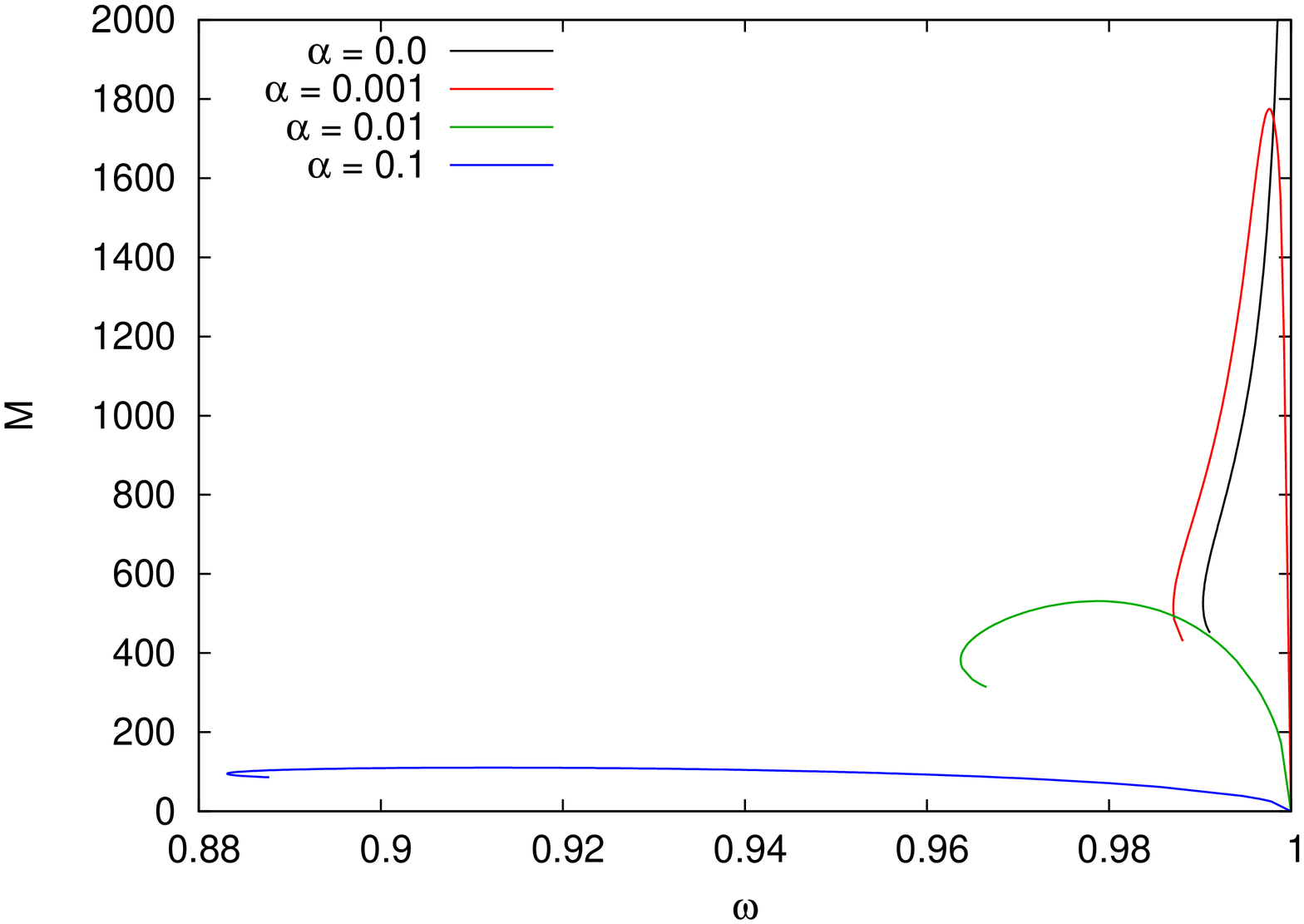}}
{\includegraphics[width=5cm, angle = 270]{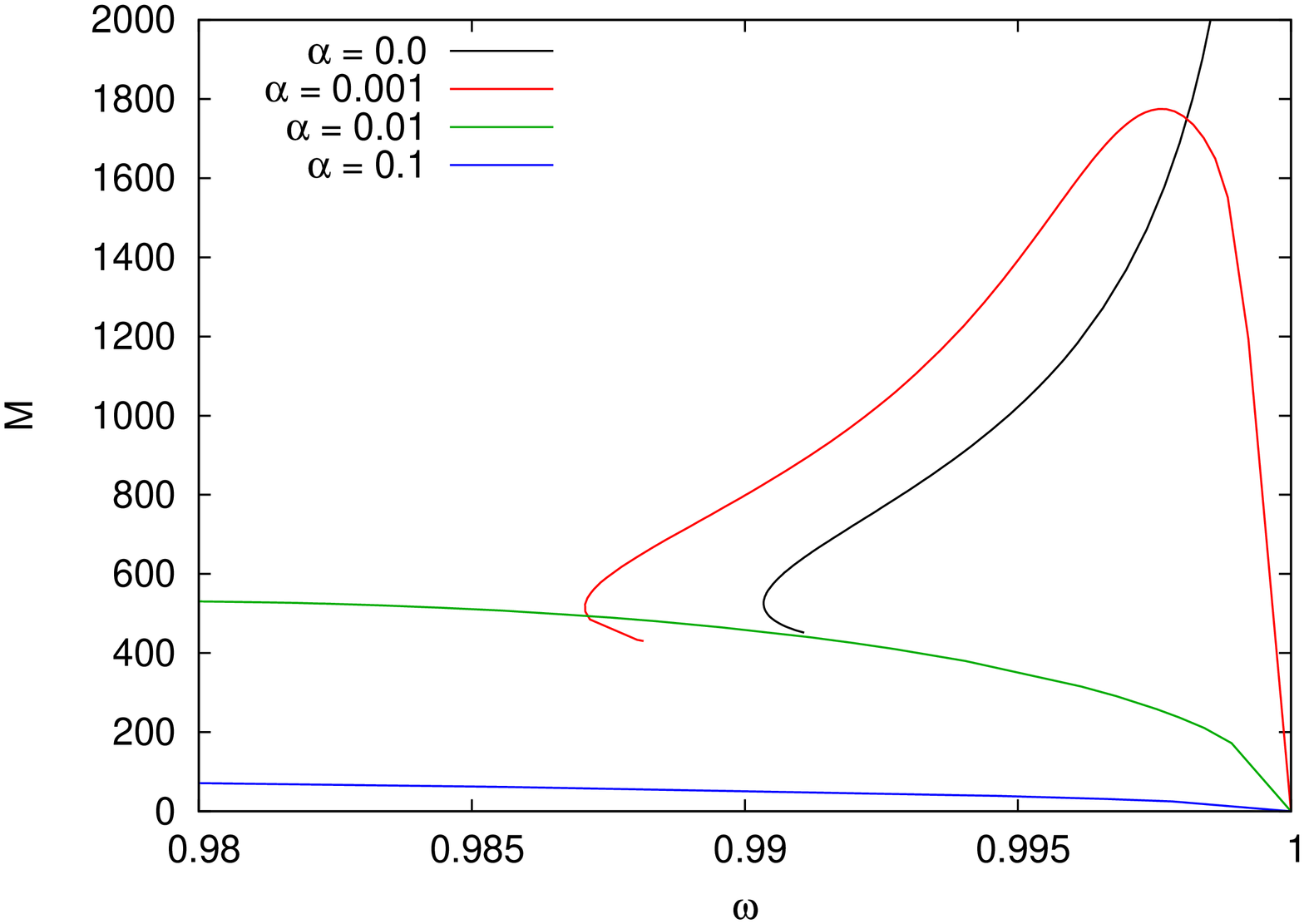}}
\caption{Left: Dependence of the mass $M$ (in units of $m/|\lambda|$) on $\omega$  for several values of $\alpha$ and $\lambda = 1$.
Right: Zoom of the region $\omega \simeq 1.0$.
\label{grav}
}
\end{center}
\end{figure}

These solutions
can be considered as continuous deformations of the Proca-Q-balls existing in the probe limit.
A common feature of the gravitating and non-gravitating cases is that both types of solutions exist for $\omega < 1$.
In the limit $\omega \to 1$ the vacuum $A_{\mu}=0$ is  approached, in particular   the mass and the charge
of the gravitating solutions approach zero. This differs from
the case $\alpha =0$  where  $M$ and $Q$ remain finite for $\omega \to 1$ (see section \ref{nu0}).
The dependence of the mass on the frequency $\omega$ is shown on Fig. \ref{grav}
for several values of the gravitating parameter $\alpha$.
The curve for $\alpha = 0$  is not continuously approached at $\omega =1$
by the curves corresponding to $\alpha > 0$. Note that for a given $\lambda$, fixing the value of $\alpha$ fixes $m$ as $m=\sqrt{2|\lambda| \alpha}\; m_{Pl}$ - see Eq. (\ref{RescPar}). Therefore, the curves in Fig. \ref{grav} correspond to typical mass values of $40 m_{Pl}$ in the $\alpha= 0.1$ curve, to a maximal value of about $80 m_{Pl}$ for  $\alpha= 0.001$.
Equivalently, the masses of the objects constructed above may be expressed as   $M=8\pi \bar{{\cal M}}(\infty) m_{Pl}^2/m$ where $\bar{{\cal M}}(x)$ is the dimensionless mass function $\bar{{\cal M}}(x)=m{\cal M}(x)$ - see Eq. (\ref{gravity_eq}). $\bar{{\cal M}}(\infty)$ is of order unity for  $\alpha \simeq1$ and decreases slowly as  $1/\alpha$ increases ($m$ decreases). So, for small enough values of $m$, large Q-star masses are possible, much similar as in the scalar case.

The numerical results exhibit the following features~:
\begin{itemize}
\item (i) The solutions exist for $\omega \in [\omega_s, 1]$ where $\omega_s$ decreases while $\alpha$ increases.
\item (ii) A second branch of solutions systematically exists in
a smaller domain of the  frequencies, say for
$\omega \in [\omega_s, \omega_{cr}]$ with $\omega_{cr} \leq 1$.
\item (iii) For $\alpha > 1$ we have $\omega_s \approx 0.815$.
\item (iv) In the limit $\omega \to \omega_{cr}$, the solution approaches a configuration
where  the $a_1$ component of the Proca field bounces at a specific radius $x_c$.
This is illustrated by Fig.\ref{grav_2}-Left.
\end{itemize}
\subsection{The case $\lambda = -1$}
\begin{figure}[t!]
\begin{center}
{\includegraphics[width=4.5cm, angle = 270]{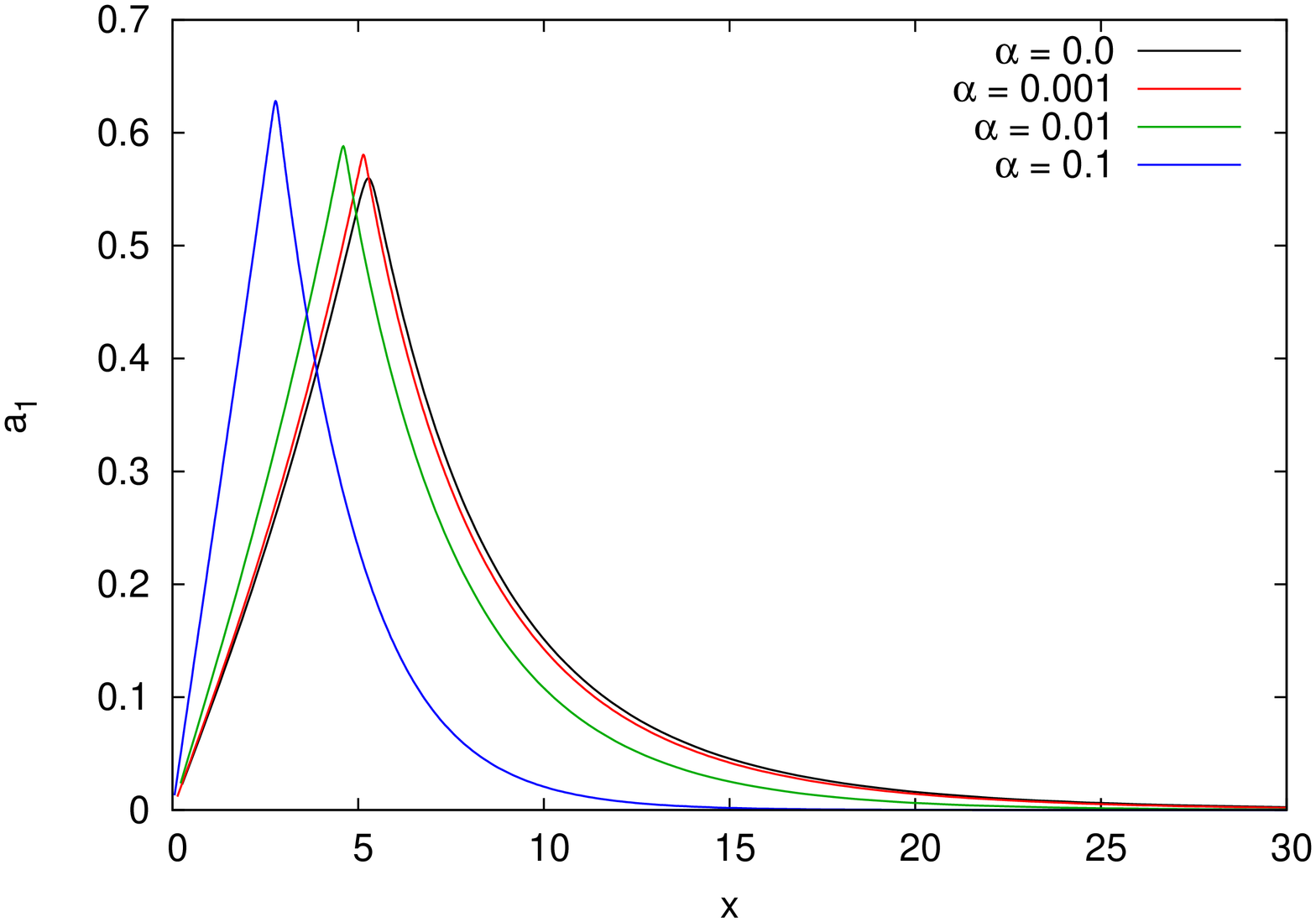}}
{\includegraphics[width=4.5cm, angle = 270]{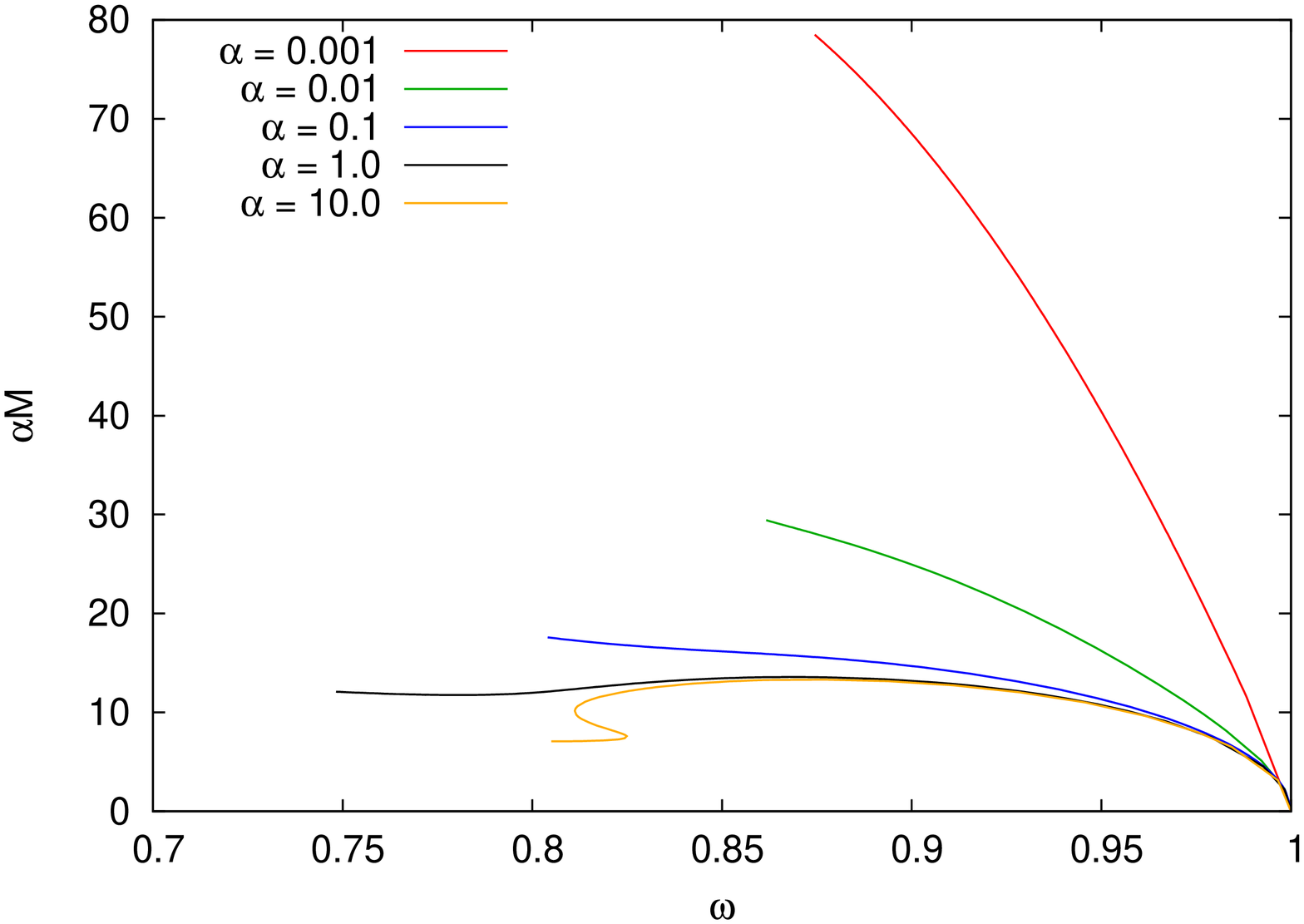}}
\caption{Left: Profiles of the bouncing $a_1$ solutions reached at $\omega \simeq \omega_{cr}$
for several values of $\alpha$.
Right: Dependence of the mass $M$ (in units of $m/|\lambda|$) on $\omega$  for several values of $\alpha$ and $\lambda = -1$. Note that the vertical axis is $\alpha M$ for better separation between the curves.
\label{grav_2}
}
\end{center}
\end{figure}

The solutions constructed with $\lambda=-1$,
also exist for $\omega \in [\omega_s,1]$ and the Proca field  tends to zero in the limit
$\omega \to 1$.
The mass as a function of $\omega$  is shown
 in Fig.\ref{grav_2}-Right for a few values of $\alpha$.
 The pattern of solutions are further characterized by the following  features~:
\begin{itemize}
\item (i) The value $\omega_s$ decreases to zero for  $\alpha \to 0$. In particular
the solutions under consideration  do not persist in the  probe limit.
\item (ii) For $\alpha \leq 1$ only  one solution exists for each value of $\omega$.
\item (iii) In the limit $\omega \to \omega_s$ a singular configuration is approached.
In particular, the maximal value of  $|a_1''(x)|$ tends rapidly to infinity.
\item (iii) For $\alpha > 1$, new branches progressively develop on specific (and small) intervals
of the frequency. On  Fig. \ref{grav_2} this appears  for the $\alpha = 10$--line.
\item (iv) Increasing $\alpha$, the  $\omega,M$ relation progressively approaches a spiral shape found by Brito et al (see Fig. 1  of \cite{BritoEtAl2015}).
\end{itemize}
\begin{figure}[b!]
\begin{center}
{\includegraphics[width=5cm, angle = 270]{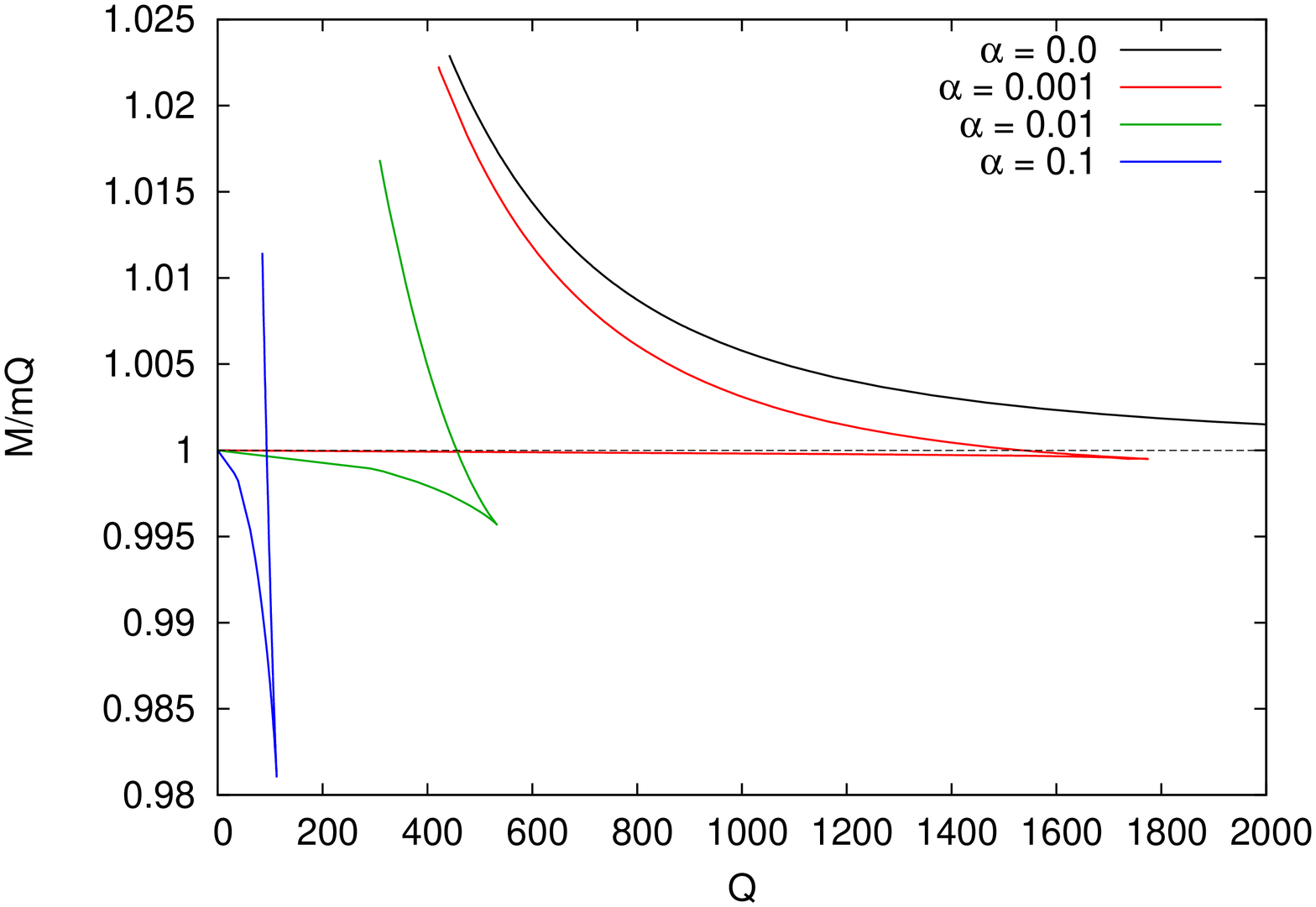}}
{\includegraphics[width=5cm, angle = 270]{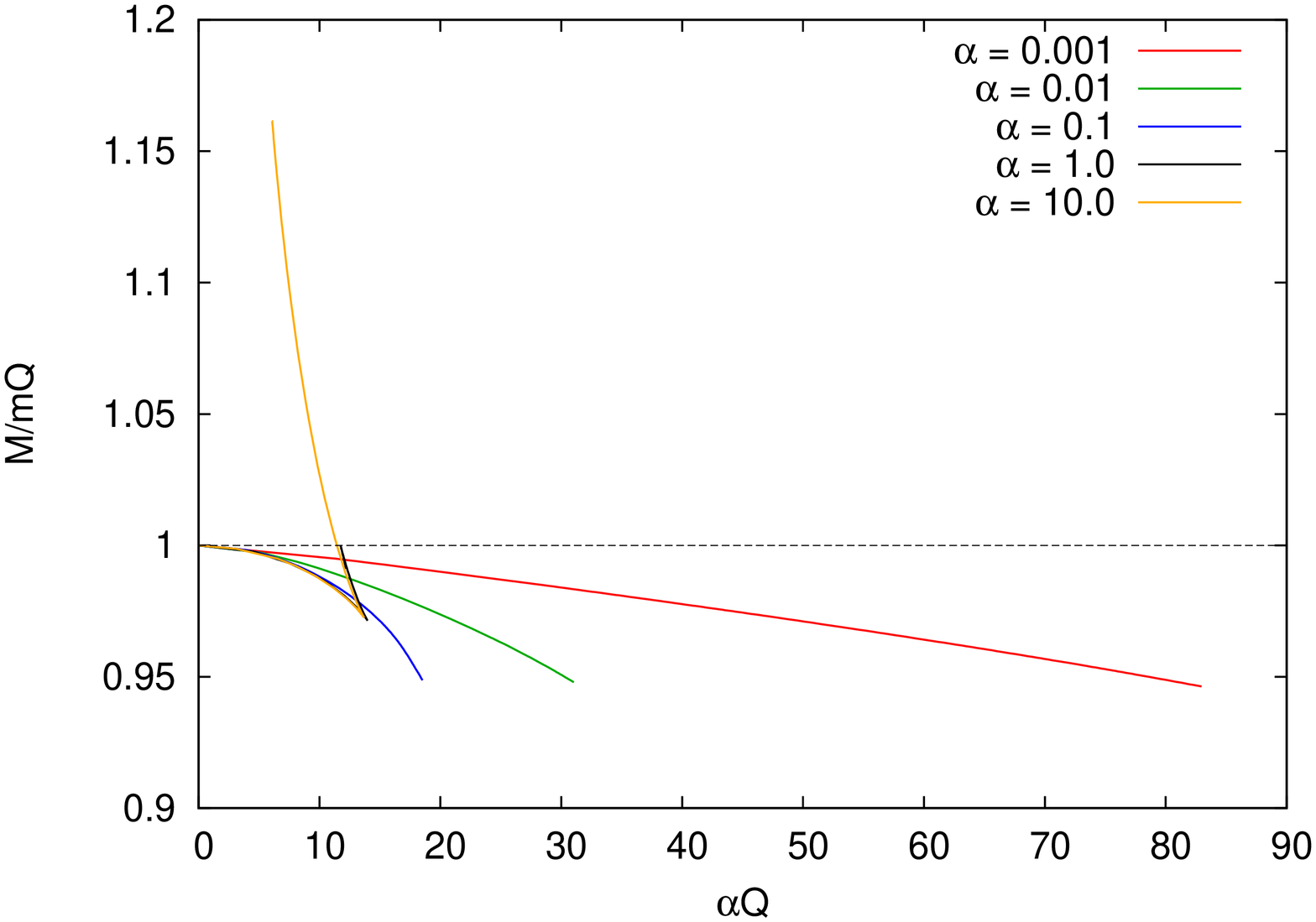}}
\caption{Dependence of the stability ratio $M/mQ$ on $Q$ for several values of $\alpha$ for $\lambda=1$ (left side)
and $\lambda=-1$ (right side). Note that the horizontal axis at the right is $\alpha Q$.
\label{grav_3}
}
\end{center}
\end{figure}
\subsection{Classical stability}
As far as the classical stability is concerned, it turns out that gravitating Proca stars
constructed above are classically stable on some part of their  domain of existence.
Fig.  \ref{grav_3} shows the ratio $M/(mQ)$ as a function of $Q$
 for a few values of $\alpha$ for $\lambda =1$ (Left) and $\lambda = -1$ (Right).
For $\lambda =1$ the plot reveals that for $\alpha > 0$ all branches systematically consist of two segments connected by a spike. The first segment, connected to
the vacuum with $Q = 0$ is systematically stable, while only a part of the second segment is stable
(see Fig.\ref{grav_3}-Left).
For $\lambda < 0$, we get only one solution with a given value of $Q$ for the small values of $\alpha$,
forming a single segment (or main segment) joining the vacuum in the limit $Q \to 0$. For $\alpha \gtrsim 1$
another segment develops, forming a spike with the main segment. For a given $Q$, the binding energy
of the second segment is smaller that the one of the main segment and eventually the solution becomes unstable.


The interval of classical stability is rather small for the  small values of $\alpha$
and  roughly coincide with the interval $[\omega_s, 1]$ for larger $\alpha$. This could be expected
since the attractive gravity naturally add binding to the elementary constituents of the soliton.

\section{Conclusion}
In this paper we have constructed a large family of vector field solitons
bounded by an appropriate potential and  eventually  by gravity. After an appropriate rescaling,
the model was found to depend on two independent coupling constants: the mass of the vector field (entering in $\alpha$) and the strength of the sextic interaction (entering through $\mu$). In addition there is a discrete parameter which we took to be the sign of the quartic term  $\lambda = \pm 1$ (after rescaling). In distinction from previous work, we found flat space solutions (although unstable) also with $\nu=0$ and even solutions with negative $\lambda$ which exist as self-gravitating solitons that are stable in quite a large range region of parameter space.

In the probe limit ($\alpha=0$) regular solutions exist for $\lambda = +1$ and $\nu \geq 0$; in the limit
$\nu = 1$, they coincide with the solutions studied in \cite{Loginov2015}.
Gravitating solutions exist for $\lambda = \pm 1$, setting $\nu=0$ the two branches reach the solution
of \cite{BritoEtAl2015} in the limit $\alpha \to \infty$.

The question how realistic these structures are, viewing the complex vector field as a possible dark matter component, requires a full sequel analysis yielding (among other things) bounds on the free parameters of the theory. For presenting a concrete example we notice that picking a value for $\alpha$ (and $\lambda$) fixes the value of the Proca mass $m$ since $2|\lambda | \alpha$ is actually the same as $m^2/m_{Pl}^2$. So e.g., $\alpha = 0.01$ with $\lambda=1/2$ correspond to a Proca mass $m = 0.1 m_{Pl}$.
The masses of the corresponding curve in Fig. \ref{grav} get therefore a maximum at about 600 units, that is about $50m_{Pl}$. As is obvious from Fig. \ref{grav}, the maximum of the dimensionless mass increases rapidly while $\alpha$ decreases, so for sufficiently small $\alpha$, stellar mass scales can be achieved too. Since these solutions are electrically neutral, they are free of constraints like the  Jetzer-Liljenberg-Skagerstam  instability \cite{JetzerEtAl1993}.
\\
\\
{\bf Acknowledgments:} Y. B. gratefully acknowledges discussions with E. Radu.
\\

\section{Appendix: comparison with boson stars}

First we write the field equations obtained after the rescaling that we used in section \ref{Rescaling}.
The gravitational equations (\ref{gravity_eq}) now take the form
\begin{eqnarray}
\label{gravity_eq_scaled}
     {\cal M}' &=& \alpha x^2 T_0^0 \ , \ \  T_0^0 = \frac{(a_0'-\omega a_1)^2}{2 A^2} + (1 + \epsilon \psi + \nu \psi^2)\frac{a_0^2}{N A^2} -  (\frac{1}{2} \psi + \frac{\epsilon }{4} \psi^2 + \frac{\nu}{6} \psi^3 )  \\
     A' &=& \frac{\alpha}{2} x A \left(\frac{a_0^2}{N^2 A^2} + a_1^2 \right) \left(1 + \epsilon \psi + \nu \psi^2 \right)
\end{eqnarray}
where $\epsilon = \pm 1 = {\rm sign}( \lambda)$ and $\psi = a_0^2/(N A^2) - N a_1^2$.
The rescaled Proca equations are obtained  along the same lines from (\ref{ProcaEqspher-0}):
\begin{eqnarray}\label{ProcaEqspher-0-Dimless}
 \omega \left( a'_0 - \omega a_1 \right) + N A^2 \left( 1 + \epsilon\psi + \nu \psi^2 \right)a_1 =0 \\
 \label{ProcaEqspher-1-Dimless}
 \frac{A N}{x^2}\left[\frac{x^2}{A} \left( a'_0 - \omega a_1 \right) \right]' - \left( 1 + \epsilon\psi + \nu \psi^2 \right)a_0 =0
\end{eqnarray}

Although the matter content is quite different, Proca stars possess several qualitative properties in common with boson  stars build out of complex scalar fields.
For the sake of comparison, we take a single complex scalar field $\Phi$ with a similar self-interaction potential:
\be \label{scalarpotential}
U(\Phi) = \frac{m^2}{2} |\Phi|^2 +  \frac{\lambda}{4} |\Phi|^4 + \frac{\nu}{6} |\Phi|^6
\ee
Boson star solutions are obtained when the scalar field $\Phi$
is parametrized according to $\Phi(x) =  \phi(r) \exp(i \omega t)$, the rescaled field equations read
\begin{eqnarray}
\label{gravity_eq_scaled_boson}
     {\cal M}' &=& \alpha x^2 T_0^0 \ , \ \  T_0^0 = N (\phi')^2 + \frac{\omega^2 \phi^2}{A^2 N} + \frac{\phi^2}{2}
      + \epsilon \frac{\phi^4}{4}+ \nu \frac{\phi^6}{6}, \ \  \\
     A' &=& 2 \alpha x \left( A (\phi')^2 + \frac{\omega^2 \phi^2}{A N^2} \right) , \\
     (x^2 A N \phi')' &=& x^2 A \left(\frac{1}{2}(\phi + \epsilon \phi^3 + \nu \phi^5) - \frac{\omega^2 \phi}{N A^2} \right)
\end{eqnarray}
The rescaling used is similar to the one of section \ref{Rescaling}.
A comparison of the spectrum of Proca stars and boson stars (both with a mass term only) is reported in
Fig. 1 of Ref. \cite{HerdeiroEtAl2016} and we do not present a similar plot here.
We  just mention a few qualitative differences~: Proca stars (resp. boson stars) exist for
$\omega \in [0.817,1.0]$ (resp. $\omega \in [0.635,1]$). Denoting $\rho_0 \equiv (M_{BS}/m_S)_{max}$
and $\rho_1 \equiv (M_{PS}/m_P)_{max}$ as the maximal mass of scalar and Proca stars respectively
(normalized as usual by the mass $m_{S,P}$ of the elementary field), one finds numerically
$\rho_P/\rho_S \approx 1.2$.
\\

\end{document}